\pdfoutput=1
\documentclass[11pt,a4paper]{article}
\usepackage[utf8]{inputenc}
\usepackage[T1]{fontenc}
\usepackage{amsmath,amssymb}
\usepackage{graphicx}
\usepackage{hyperref}
\usepackage{xcolor}
\usepackage{listings}
\usepackage{booktabs}
\usepackage{enumitem}
\usepackage{geometry}
\usepackage{fancyhdr}
\usepackage{titlesec}
\usepackage{float}
\usepackage{caption}
\usepackage{subcaption}
\usepackage{tikz}
\usetikzlibrary{shapes.geometric, arrows.meta, positioning, fit, backgrounds, calc, decorations.pathreplacing}
\usepackage{framed}
\newcounter{algline}

\hypersetup{colorlinks=true,linkcolor=blue!60!black,citecolor=blue!60!black,urlcolor=blue!60!black}

\lstdefinestyle{json}{
  basicstyle=\small\ttfamily,
  frame=single,
  backgroundcolor=\color{gray!8},
  breaklines=true,
  columns=fullflexible,
  keepspaces=true,
  showstringspaces=false,
  numbers=none,
  xleftmargin=0.5em,
  xrightmargin=0.5em,
  aboveskip=0.8em,
  belowskip=0.8em,
}

\title{\textbf{Bridging Protocol and Production: Design Patterns for\\Deploying AI Agents with Model Context Protocol}}

\author{
  Vasundra Srinivasan\\
  AI Architect\\
  \small{Author, \textit{Data Engineering for Multimodal AI} (O'Reilly)}
}

\date{March 2026}

\begin{document}
\maketitle

\noindent\fbox{\parbox{\dimexpr\textwidth-2\fboxsep-2\fboxrule}{%
\small\textbf{Disclaimer.} This paper represents the author's independent research and personal views, conducted entirely outside the scope of any employment or contractual obligation. It is not sponsored by, endorsed by, affiliated with, or authorized by the author's employer, any client organization, or any technology vendor referenced herein. The author received no funding, compensation, or resources from any organization for this work. No proprietary, confidential, trade-secret, or non-public information is disclosed; all technical observations are derived solely from the author's general professional experience with publicly available protocols, open-source tools, and published specifications. All platform vendor and client organization names have been redacted to preserve confidentiality. The ``enterprise AI agent platform'' referenced throughout is used as a generic descriptor and does not identify any specific commercial product.}}

\begin{abstract}
The Model Context Protocol (MCP) standardizes how AI agents discover and invoke external tools, with over 10,000 active servers and 97 million monthly SDK downloads as of early 2026. Yet MCP does not yet standardize how agents \textit{safely operate} those tools at production scale. Three protocol-level primitives remain missing: identity propagation, adaptive tool budgeting, and structured error semantics. This paper identifies these gaps through field lessons from an enterprise deployment of an AI agent platform integrated with a major cloud provider's MCP servers (client name redacted). We propose three mechanisms to fill them: (1) the \textit{Context-Aware Broker Protocol} (CABP), which extends JSON-RPC with identity-scoped request routing via a six-stage broker pipeline; (2) \textit{Adaptive Timeout Budget Allocation} (ATBA), which frames sequential tool invocation as a budget allocation problem over heterogeneous latency distributions; and (3) the \textit{Structured Error Recovery Framework} (SERF), which provides machine-readable failure semantics that enable deterministic agent self-correction. We organize production failure modes into five design dimensions (server contracts, user context, timeouts, errors, and observability), document concrete failure vignettes, and present a production readiness checklist. All three algorithms are formalized as testable hypotheses with reproducible experimental methodology. Field observations demonstrate that while MCP provides a solid protocol foundation, reliable agent tool integration requires infrastructure-level mechanisms that the specification does not yet address.
\end{abstract}

\section{Introduction}

The Model Context Protocol (MCP) \cite{mcp-spec}, introduced by Anthropic in November 2024 and donated to the Linux Foundation's Agentic AI Foundation in December 2025, provides a standardized JSON-RPC 2.0-based protocol for AI agents to discover and invoke external tools. In its first year, the ecosystem has grown rapidly: as of early 2026, over 10,000 active MCP servers are in production, more than 500 MCP clients exist across major platforms (Claude, ChatGPT, Cursor, VS Code, Replit), and monthly SDK downloads have reached 97 million \cite{anthropic-foundation}.

A natural question arises: why adopt MCP rather than exposing traditional REST APIs? The answer lies in three capabilities that REST alone does not provide. First, MCP's \texttt{tools/list} endpoint enables runtime tool discovery. Agents can learn what tools are available without being hard-coded to specific endpoints. Second, the \texttt{initialize} handshake provides capability negotiation, allowing client and server to agree on protocol features (such as async task support) before any tool is invoked. Third, MCP is converging as the standard supported by every major AI platform; building a REST adapter for each platform is more costly than building one MCP server that all platforms can consume. In this sense, MCP functions as a universal adapter layer for agent-to-tool communication.

While MCP standardizes how agents call tools, it does not yet standardize how agents safely operate them. MCP's simplicity gets you to a demo in 30 minutes. That same simplicity will break your production deployment if you don't account for what the protocol leaves out. Three primitives are missing at the protocol level: identity propagation (who is this request for?), adaptive tool budgeting (how long should each tool get?), and structured error semantics (what should the agent do when a tool fails?). This paper documents the lessons learned from confronting those gaps in production.

The deployment involved an enterprise AI agent platform with native MCP client support as an employee-facing agent that integrates with a major cloud provider's APIs via an MCP server (the client organization's name is redacted for confidentiality). The agent handles cloud resource usage-limit management, a multi-step workflow requiring sequential tool calls to fetch projects, services, resource limits, and submit limit increase requests. This deployment exposed fundamental design challenges that are common to any enterprise MCP integration but are under-documented in the existing literature.

Our contributions include the following.

\begin{enumerate}[leftmargin=*]
  \item \textbf{A systematic taxonomy of production MCP failure modes}, organized into five design dimensions. server contracts, user context, timeouts, errors, and observability.
  \item \textbf{The Context-Aware Broker Protocol (CABP)} for MCP user context propagation, formalizing the broker/gateway pattern with verifiable security properties.
  \item \textbf{The Adaptive Timeout Budget Allocation (ATBA) algorithm} for dynamic distribution of planner timeout budgets across sequential tool chains.
  \item \textbf{The Structured Error Recovery Framework (SERF)} enabling deterministic agent self-correction through machine-readable error taxonomies.
  \item \textbf{Practical guidance and a production readiness checklist} grounded in a real deployment with specific failure examples and resolution patterns.
\end{enumerate}

\noindent While LLM reasoning and planning capabilities have advanced rapidly, the reliability of agent tool use remains constrained by the surrounding infrastructure. Recent work has addressed production agentic AI workflows at a general architectural level \cite{agentic-ai-guide}, but MCP introduces protocol-specific concerns (JSON-RPC identity propagation, capability negotiation, turn-budget allocation) that do not arise in traditional REST-based agent architectures. We also examine broker-layer security threats, including prompt injection via tool responses, data exfiltration, privilege escalation through tool chaining, and denial of service via resource exhaustion. This paper argues that identity propagation, timeout budgeting, and structured failure semantics are necessary protocol-level primitives for reliable agent ecosystems.

\section{Background and Related Work}

\subsection{The Model Context Protocol}

MCP defines a client-server protocol over Streamable HTTP using JSON-RPC 2.0 \cite{jsonrpc}. The protocol supports tool discovery (\texttt{tools/list}), tool invocation (\texttt{tools/call}), resource access, and prompt templates. The current specification version (2025-11-25) includes experimental support for asynchronous task execution via MCP Tasks \cite{mcp-spec}.

The protocol lifecycle begins with an \texttt{initialize} handshake where client and server negotiate the protocol version and advertise supported capabilities. Capability negotiation matters: a server advertising \texttt{experimental.tasks: true} enables asynchronous operations. Omit that flag, and all operations become synchronous and subject to timeout constraints.

\subsection{The AI Agent Platform}

The deployment uses an enterprise AI agent platform with native MCP client support. The platform's planning engine orchestrates agent behavior using a concurrent multi-agent orchestration pattern. The agent planner selects tools based on their name and description as returned by \texttt{tools/list}. It does not inspect implementation details. The tool metadata is, in practice, the API contract between server and agent.

The agent planner operates within a turn budget. Each agent turn has a timeout window within which all tool calls must complete. Sequential tool chains (where each call depends on the previous result) consume this budget additively. If any tool in the chain exceeds its individual timeout, or if the cumulative chain exceeds the session timeout, the entire turn fails.

\subsection{Why MCP? A Protocol Comparison}

Consider why MCP over alternatives. Table~\ref{tab:protocol-comparison} summarizes the trade-offs between MCP, REST, and GraphQL for agent-to-tool communication.

\begin{table}[H]
\centering
\caption{Protocol comparison for agent-to-tool communication}
\label{tab:protocol-comparison}
\small
\begin{tabular}{p{2.8cm}p{3.6cm}p{3.6cm}p{3.6cm}}
\toprule
\textbf{Dimension} & \textbf{MCP} & \textbf{REST} & \textbf{GraphQL} \\
\midrule
Tool discovery & Runtime via \texttt{tools/list}; agent learns available tools dynamically & None; endpoints must be hard-coded or documented out-of-band & Introspection query provides schema, but no semantic descriptions \\
\addlinespace
Capability negotiation & \texttt{initialize} handshake advertises features (async, sampling, etc.) & None; client must know server capabilities a priori & None; schema is static \\
\addlinespace
Agent compatibility & Supported natively by Claude, ChatGPT, Cursor, VS Code, Copilot, and 500+ clients & Requires per-platform adapter & Requires per-platform adapter \\
\addlinespace
Async operations & MCP Tasks (experimental) with polling and progress & Webhooks or polling (no standard) & Subscriptions (WebSocket) \\
\addlinespace
Error semantics & Two-tier: protocol (JSON-RPC) + tool (\texttt{isError}) & HTTP status codes only & \texttt{errors} array with extensions \\
\addlinespace
User context & Gap: not in spec (broker pattern required) & Standard: OAuth headers & Standard: context headers \\
\addlinespace
Ecosystem momentum & 10K+ servers, 97M SDK downloads/month, Linux Foundation governance & Mature, universal & Mature, strong in web/mobile \\
\bottomrule
\end{tabular}
\end{table}

MCP's primary advantage is agent-native integration: a single MCP server is consumable by every major AI platform without per-platform adapter code. Its primary disadvantage is the user context gap that this paper addresses. REST and GraphQL are mature but require custom agent-integration layers for each platform.

\subsection{Related Work}

API gateway patterns exist in traditional infrastructure \cite{kong-ai-gateway, envoy-proxy}. Applying them to MCP-specific concerns (tool-level ACLs, JSON-RPC context injection, response sanitization) requires new thinking.

Several recent works have examined MCP from complementary perspectives. Hou et al.\ \cite{hou-mcp-landscape} provide the most comprehensive MCP security taxonomy to date, identifying 16 threat scenarios across four attacker types mapped to the MCP lifecycle. Their analysis is broad and taxonomic; our work contributes concrete broker-layer mitigations grounded in a production deployment. Hasan et al.\ \cite{hasan-mcp-glance} conduct an empirical study of 1,899 open-source MCP servers, finding that 7.2\% contain general vulnerabilities and 5.5\% exhibit tool poisoning risks. The MCPTox benchmark \cite{mcptox} further quantifies tool poisoning attack success rates across 45 real-world MCP servers with 353 tools. These empirical findings validate our emphasis on tool description integrity and broker-layer response sanitization.

Li et al.\ \cite{li-ca-mcp} propose Context-Aware MCP (CA-MCP), which addresses context propagation through a shared memory store accessible to multiple agents. Our CABP uses a different mechanism: rather than shared state, CABP injects identity context from JWT claims into individual JSON-RPC requests at the broker layer. This preserves statelessness and aligns with enterprise zero-trust security models. Both approaches have value. CA-MCP handles collaborative multi-agent memory; CABP handles identity-scoped access control. Krishnan \cite{krishnan-multi-agent} examines multi-agent MCP architectures for enterprise integration, providing additional context for how sequential tool chains interact across agent boundaries.

Dhar et al.\ \cite{agentic-ai-guide} present a practical guide for production-grade agentic AI workflows, covering containerized deployment, tool-first design, and operational resilience. Their work provides a broad architectural treatment; our paper contributes MCP-specific design patterns (CABP, ATBA, SERF) with formal definitions and a concrete deployment case study, addressing protocol-level concerns (timeout budget allocation, structured error recovery, capability negotiation) that are outside the scope of their general-purpose framework.

Industry reports from Zuplo \cite{zuplo-mcp-report} and Astrix Security \cite{astrix-mcp-security} have documented authentication challenges and credential management practices across the MCP ecosystem. Astrix's analysis of 5,200 MCP servers found that 53\% rely on insecure long-lived static secrets, while only 8.5\% use OAuth. This underscores the urgency of the broker-layer security patterns we propose.

\section{System Architecture}

\subsection{Deployment Overview}

The deployment consists of four layers. 
\begin{enumerate}[leftmargin=*]
  \item \textbf{Support Interface}: Support engineers initiate agent interactions from service tickets via the platform's chat interface.
  \item \textbf{AI Agent (Planner)}: Orchestrates the workflow, managing intents (``Retrieve User Identity,'' ``Submit Limit Increase''), session state, and sequential MCP tool invocations.
  \item \textbf{MCP Server}: Exposes four tools over JSON-RPC 2.0: \texttt{FetchResources}, \texttt{FetchServices}, \texttt{FetchUsageLimits}, and \texttt{CreateLimitRequest}. Stateless, with no authentication in the proof-of-concept phase.
  \item \textbf{Cloud Provider APIs}: Downstream services for resource management, service discovery, usage limits, and identity/access management.
\end{enumerate}

\begin{figure}[H]
\centering
\resizebox{\textwidth}{!}{%
\begin{tikzpicture}[
  node distance=0.6cm and 1.2cm,
  box/.style={rectangle, draw, rounded corners=3pt, minimum width=2.8cm, minimum height=0.9cm, align=center, font=\small},
  layer/.style={rectangle, draw, dashed, rounded corners=5pt, inner sep=8pt, fill=#1},
  arr/.style={-{Stealth[length=6pt]}, thick},
  darr/.style={-{Stealth[length=6pt]}, thick, dashed},
]
\node[box, fill=blue!8] (ui) {Support\\Interface};
\node[box, fill=blue!15, right=1.5cm of ui] (planner) {Agent\\Planner};
\node[box, fill=orange!15, right=1.5cm of planner] (broker) {Broker /\\Gateway};
\node[box, fill=green!12, right=1.5cm of broker] (mcp) {MCP\\Server};
\node[box, fill=gray!12, right=1.5cm of mcp] (cloud) {Cloud\\Provider APIs};

\draw[arr] (ui) -- node[above, font=\scriptsize] {chat} (planner);
\draw[arr] (planner) -- node[above, font=\scriptsize] {JSON-RPC} (broker);
\draw[arr] (broker) -- node[above, font=\scriptsize] {enriched} node[below, font=\scriptsize] {JSON-RPC} (mcp);
\draw[arr] (mcp) -- node[above, font=\scriptsize] {REST/} node[below, font=\scriptsize] {gRPC} (cloud);

\node[below=0.4cm of ui, font=\scriptsize\itshape, text=gray] {Service ticket};
\node[below=0.4cm of planner, font=\scriptsize\itshape, text=gray] {Tool selection};
\node[below=0.4cm of broker, font=\scriptsize\itshape, text=gray] {JWT + ACL};
\node[below=0.4cm of mcp, font=\scriptsize\itshape, text=gray] {4 tools};
\node[below=0.4cm of cloud, font=\scriptsize\itshape, text=gray] {Downstream};

\draw[darr, blue!60] ([yshift=-1.1cm]ui.south) -- node[below, font=\scriptsize, text=blue!60] {OpenTelemetry trace propagation} ([yshift=-1.1cm]cloud.south);

\end{tikzpicture}%
}
\caption{End-to-end deployment architecture. Solid arrows indicate request flow; the dashed arrow shows OpenTelemetry trace propagation across all layers.}
\label{fig:architecture}
\end{figure}
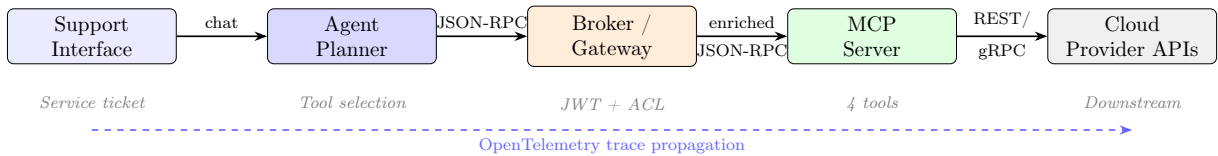

The agent executes a sequential chain. It first fetches the user's cloud projects, then the services within a selected project, then the usage limits for a selected service, and finally submits a limit increase request. Each step depends on the output of the previous step, making the chain inherently sequential and sensitive to per-tool latency. Figure~\ref{fig:architecture} illustrates the end-to-end deployment architecture.

\subsection{Tool Interface Design}

Each MCP tool is defined with typed JSON Schema inputs (distinguishing required from optional parameters), structured JSON outputs with field descriptions, explicit documentation of side effects (read-only vs. write), and idempotency guarantees. An example tool definition:

\begin{lstlisting}[style=json]
{
  "name": "FetchUsageLimits",
  "description": "Returns current usage limits and
    consumption for a specified cloud service within
    a project. Read-only. Idempotent. Safe to retry.",
  "inputSchema": {
    "type": "object",
    "properties": {
      "project_id": { "type": "string", "description":
        "Cloud project identifier" },
      "service_name": { "type": "string", "description":
        "Fully qualified service name" }
    },
    "required": ["project_id", "service_name"]
  }
}
\end{lstlisting}

This level of detail functions as the actual interface the agent planner uses when deciding whether and how to invoke a tool.

\section{Design Dimension 1: Server Contract Design}

\subsection{The Problem}

The agent planner selects tools from the \texttt{tools/list} response based solely on tool names and descriptions. Unlike traditional API clients programmed to call specific endpoints, the agent must reason about which tool to use. The \texttt{tools/list} response becomes the most consequential artifact an MCP server ships.

\subsection{Failure Mode: Ambiguous Tool Descriptions}

In early iterations, the agent planner was observed failing to select the correct tool when descriptions were vague. A tool named \texttt{get\_data\_v2} with description ``Gets data'' provides insufficient signal for tool selection. The agent either skipped the tool entirely or invoked it with incorrect parameters, leading to fabricated responses.

\subsection{Recommendations}

\begin{enumerate}[leftmargin=*]
  \item \textbf{Use descriptive, action-oriented tool names} that convey the operation (e.g., \texttt{FetchUsageLimits} rather than \texttt{get\_data}).
  \item \textbf{Write descriptions as if they are API documentation}: what the tool does, when to call it, what it returns, and what side effects it has.
  \item \textbf{Provide typed JSON Schema inputs} with required/optional distinctions and field-level descriptions.
  \item \textbf{Document idempotency guarantees} explicitly. The agent may retry tools, and knowing whether a retry is safe affects the agent's recovery strategy.
\end{enumerate}

A practical concern is description length. The MCP specification imposes no character limit on tool descriptions, but the agent planner's context window has a finite budget. In field observations, two to four sentences per tool strikes the right balance: enough to convey what the tool does, when to call it, what it returns, and what side effects it has (or explicitly that it has none), but not so verbose that the agent is overwhelmed by irrelevant detail. Descriptions that are too short force the agent to guess; descriptions that are too long dilute the signal among noise.

\subsection{Tool Versioning}

A related challenge is tool versioning. Appending version suffixes to tool names (e.g., \texttt{FetchResources\_v2}) is an anti-pattern: the agent cannot reliably determine which version to invoke. Instead, we recommend evolving tools by adding optional fields to inputs (preserving backward compatibility), adding new fields to outputs (never removing existing ones), and using capability negotiation to advertise new features. If a breaking change is unavoidable, the correct approach is to deploy a new server version and update the client's MCP registry entry, rather than shipping two incompatible tool names that compete for the agent's attention.

\subsection{Capability Negotiation}

The \texttt{initialize} handshake is a contract, not a formality. Servers must:

\begin{itemize}[leftmargin=*]
  \item Validate the client's requested protocol version and return an error (not silently downgrade) if incompatible.
  \item Advertise \texttt{experimental.tasks: true} if the server supports long-running operations.
  \item Reject all requests except \texttt{ping} before initialization completes, returning \texttt{-32600} (Invalid Request).
\end{itemize}

\section{Design Dimension 2: The Broker Pattern}

\subsection{The Problem}

User identity cannot currently be propagated via MCP request headers to the server. The JSON-RPC protocol does not include a standard mechanism for carrying user context (identity, permissions, tenant scope) with tool invocations. Without this, an MCP server cannot scope responses to the requesting user. Every tool returns all data for all users, or worse, data for the wrong tenant.

User context propagation remains the largest unsolved problem in the MCP specification for enterprise use.

\subsection{Approaches}

We identify three approaches to user context propagation, in order of increasing maturity.

\textbf{Approach 1: Input Parameter Workaround.} In the observed agent platform, a pre-invocation workflow captures the running user's email and stores it in a session variable. This value is passed as an explicit input parameter on every tool call. The server scopes its database queries by this parameter. This was shipped in production. It works but is fragile: it relies on every tool accepting and correctly handling the user parameter, and it provides no cryptographic verification of the user's identity.

\textbf{Approach 2: Broker/Gateway (Recommended).} A gateway service sits between the agent and the MCP server. It intercepts JSON-RPC requests, extracts claims from the OAuth token in the Authorization header (validating JWT signature, expiry, and issuer), and injects \texttt{tenant\_id}, \texttt{user\_id}, and permission scopes into the request context before forwarding to the MCP server. The server reads identity from this enriched context rather than from raw input parameters.

\textbf{Approach 3: Native Protocol Support (Future).} The MCP specification could evolve to support user context natively in the transport layer, following the OAuth Resource Server pattern with Resource Indicators. This is under discussion but has no defined timeline.

A practical concern: teams tempt themselves to defer the broker investment and wait for native protocol support. Don't. Production deployments cannot wait for a specification change with no defined timeline. The broker solves the problem today. And even after native protocol support arrives, the broker stays necessary for ACL enforcement, audit logging, response sanitization, and credential management. It's not a workaround. It's a permanent architectural component.

\subsection{Broker Architecture}

The broker must fulfill six non-negotiable responsibilities.

\begin{enumerate}[leftmargin=*]
  \item \textbf{JWT Extraction and Validation}: Parse the OAuth token from the Authorization header. Validate signature, expiry, and issuer. Extract \texttt{user\_id}, \texttt{tenant\_id}, and scopes.
  \item \textbf{Context Injection}: Inject identity claims into the JSON-RPC request context before forwarding. The MCP server reads from context, not raw parameters.
  \item \textbf{Tool-Level ACL Enforcement}: Not all users can call all tools. The broker checks whether the user's role has permission to invoke the requested tool.
  \item \textbf{Response Sanitization}: Validate that the server's response does not contain cross-tenant data. Strip fields the user should not see.
  \item \textbf{Credential Vault}: External API keys are never exposed to the agent. Credentials are stored in an encrypted vault and injected at execution time.
  \item \textbf{Audit Trail}: Log every request with identity, tool name, sanitized inputs, output summary, latency, and success/failure status.
\end{enumerate}

\noindent\textbf{Distinction from CA-MCP.} Li et al.\ \cite{li-ca-mcp} propose Context-Aware MCP (CA-MCP), which propagates context via a shared memory store accessible to multiple agents. CABP uses a different mechanism: it injects identity claims from JWT tokens into individual JSON-RPC request contexts at the broker layer, maintaining stateless request processing. This aligns with enterprise zero-trust architectures where each request is independently authenticated and authorized. CA-MCP targets collaborative multi-agent memory sharing; CABP targets identity-scoped access control within a single request.

\begin{figure}[H]
\centering
\begin{tikzpicture}[
  stage/.style={rectangle, draw, rounded corners=2pt, minimum width=1.6cm, minimum height=0.8cm, align=center, font=\scriptsize, fill=#1},
  arr/.style={-{Stealth[length=5pt]}, thick},
  node distance=0.35cm,
]
\node[stage=orange!15] (s1) {1. Token\\Extract};
\node[stage=orange!20, right=of s1] (s2) {2. Claim\\Validate};
\node[stage=yellow!20, right=of s2] (s3) {3. ACL\\Resolve};
\node[stage=blue!10, right=of s3] (s4) {4. Context\\Inject};
\node[stage=green!12, right=of s4] (s5) {5. Response\\Sanitize};
\node[stage=gray!15, right=of s5] (s6) {6. Audit\\Emit};

\draw[arr] (s1) -- (s2);
\draw[arr] (s2) -- (s3);
\draw[arr] (s3) -- (s4);
\draw[arr] (s4) -- (s5);
\draw[arr] (s5) -- (s6);

\node[left=0.5cm of s1, font=\scriptsize\itshape, text=gray, align=center] {Agent\\request};
\node[right=0.5cm of s6, font=\scriptsize\itshape, text=gray, align=center] {Audit\\log};

\draw[-{Stealth[length=4pt]}, dashed, red!60, thick] (s2.south) -- ++(0,-0.6) node[below, font=\tiny, text=red!60] {Reject: invalid/expired JWT};
\draw[-{Stealth[length=4pt]}, dashed, red!60, thick] (s3.south) -- ++(0,-0.6) node[below, font=\tiny, text=red!60] {Reject: insufficient role};

\node[below=1.4cm of s4, font=\scriptsize\itshape, text=blue!60] (fwd) {Forward enriched request to MCP Server};
\draw[-{Stealth[length=4pt]}, blue!60, thick] (s4.south) -- (fwd);
\end{tikzpicture}
\caption{The CABP broker pipeline. Stages 2 and 3 can reject requests (dashed red arrows). Stage 4 forwards the enriched request to the MCP server. Stage 5 filters the response before returning to the agent.}
\label{fig:broker-pipeline}
\end{figure}
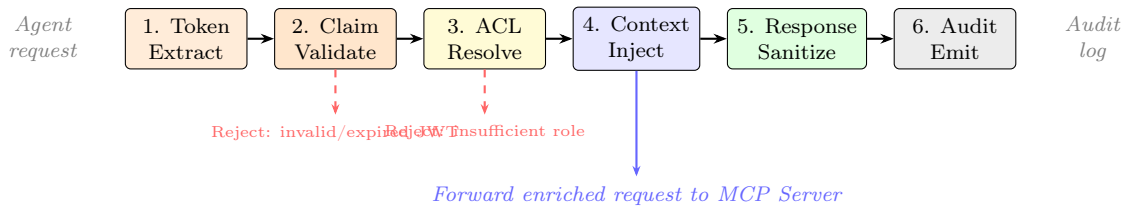

Figure~\ref{fig:broker-pipeline} illustrates the six-stage CABP pipeline that every JSON-RPC request traverses.

A few operational concerns matter in practice. Token refresh mid-session is the first. The broker must validate the JWT on every request, not just at session start. If a token expires between tool calls in a sequential chain, the broker returns \texttt{-32600} (Invalid Request) with a descriptive message, and the agent platform's OAuth integration handles the refresh and retries the call. Never cache tokens. Validate fresh on each request.

Service-to-service calls present a second issue. Not all MCP invocations originate from a human user. Automated workflows and batch jobs use service account identities with dedicated tenant scopes. The broker validates these JWTs identically but maps them to service principals rather than human users. Service accounts should receive narrower tool-level ACLs than human users. A batch processing job should not have permission to invoke write operations like \texttt{CreateLimitRequest}.

Multi-tenant data isolation requires attention. The broker injects \texttt{tenant\_id} from the JWT into every request context, and the MCP server must filter all queries by this tenant identifier without exception. Response sanitization in the broker provides a second layer of defense, validating that no cross-tenant records appear in the response payload. Track cross-tenant access attempts (target: zero; alert immediately if non-zero) and conduct periodic security audits of the broker's filtering logic.

The security review process is the final concern. Exposing internal APIs via MCP requires the same rigor as any external API exposure, plus three agent-specific considerations. Can the agent be manipulated into calling tools with malicious inputs? Validate inputs at both broker and server. Can tool responses leak sensitive data to the agent's context? Filter outputs and redact fields. Can a compromised agent escalate privileges? Enforce strict least-privilege ACLs per tool at the broker.

\subsection{Implementation Options}

Production implementations include Kong AI Gateway with the MCP Proxy plugin (providing authentication, per-tool ACLs, and rate limiting out of the box), Cloudflare Workers with Zero Trust integration, Envoy AI Gateway, and custom middleware in Flask or Express. The trade-off is flexibility versus operational overhead: Kong and Envoy provide authentication, rate limiting, and ACL enforcement as configuration rather than code, but require infrastructure investment and operational expertise. Custom middleware in Flask or Express gives full control over context injection logic and can be deployed in under 200 lines of code, but every cross-cutting concern (rate limiting, circuit breaking, audit logging) must be implemented manually. Our recommendation: start with custom middleware for proof-of-concept (1--3 servers), migrate to Kong or Envoy for production at scale (10+ servers), where the operational cost of maintaining custom middleware across many servers exceeds the setup cost of an API gateway.

\section{Design Dimension 3: Timeouts and Asynchronous Operations}

\subsection{The Problem}

The agent planner operates within a turn budget. In the deployment, a sequential chain of four MCP calls must complete within this budget:

\begin{table}[H]
\centering
\caption{Latency budget for a 4-tool sequential chain}
\begin{tabular}{lcc}
\toprule
\textbf{Tool} & \textbf{Typical Latency} & \textbf{Category} \\
\midrule
FetchResources & 200ms & Read \\
FetchServices & 150ms & Read \\
FetchUsageLimits & 300ms & Read \\
CreateLimitRequest & 400ms & Write \\
Planner overhead & $\sim$500ms & System \\
\midrule
\textbf{Total} & \textbf{$\sim$1,550ms} & \\
\bottomrule
\end{tabular}
\end{table}

While our typical chain completes well within budget, the failure mode is catastrophic when it does not. Four failure scenarios were observed:

\begin{enumerate}[leftmargin=*]
  \item \textbf{Individual tool timeout ($>$30s)}: The standard MCP client timeout triggers. The agent receives an error or empty response.
  \item \textbf{Chain timeout ($>$100s)}: The MCP session timeout (100s default) expires. The entire turn is lost.
  \item \textbf{Planner budget exhaustion}: The agent planner runs out of its turn budget. The conversation ends mid-flow.
  \item \textbf{Orphaned responses}: The tool returns data, but the agent has already moved on. The data is retrieved but never presented to the user.
\end{enumerate}

\begin{figure}[H]
\centering
\begin{tikzpicture}[x=0.09cm]
\node[anchor=east, font=\small\bfseries] at (-2, 1.2) {Nominal};
\draw[fill=gray!10, draw=gray!40] (0,0.4) rectangle (100,1.2);
\draw[fill=blue!25] (0,0.4) rectangle (13,1.2);
\draw[fill=blue!30] (13,0.4) rectangle (23,1.2);
\draw[fill=blue!35] (23,0.4) rectangle (43,1.2);
\draw[fill=orange!30] (43,0.4) rectangle (69,1.2);
\draw[fill=gray!25] (69,0.4) rectangle (85,1.2);
\node[font=\tiny, anchor=south] at (6.5,1.25) {Fetch};
\node[font=\tiny, anchor=south] at (6.5,0.95) {Resources};
\node[font=\tiny, anchor=south] at (18,1.25) {Fetch};
\node[font=\tiny, anchor=south] at (18,0.95) {Svc};
\node[font=\tiny, anchor=south] at (33,1.25) {FetchUsage};
\node[font=\tiny, anchor=south] at (33,0.95) {Limits};
\node[font=\tiny, anchor=south] at (56,1.25) {Create};
\node[font=\tiny, anchor=south] at (56,0.95) {Limit};
\node[font=\tiny, anchor=south] at (77,1.25) {Planner};
\node[font=\small, green!50!black] at (92, 0.8) {\checkmark};
\node[font=\tiny, text=gray] at (100, 0.05) {100s};

\node[anchor=east, font=\small\bfseries] at (-2, -1.3) {Failure};
\draw[fill=gray!10, draw=gray!40] (0,-2.1) rectangle (100,-1.3);
\draw[fill=blue!25] (0,-2.1) rectangle (13,-1.3);
\draw[fill=blue!30] (13,-2.1) rectangle (23,-1.3);
\draw[fill=red!35] (23,-2.1) rectangle (100,-1.3);
\node[font=\tiny, anchor=south] at (6.5,-1.25) {Fetch};
\node[font=\tiny, anchor=south] at (6.5,-1.55) {Resources};
\node[font=\tiny, anchor=south] at (18,-1.25) {Fetch};
\node[font=\tiny, anchor=south] at (18,-1.55) {Svc};
\node[font=\tiny, anchor=south] at (61.5,-1.25) {FetchUsageLimits (slow: p99 spike)};
\node[font=\small, red!70!black] at (104, -1.7) {\texttimes};
\draw[dashed, red!60, thick] (100,-2.4) -- (100,1.5);
\node[font=\tiny, red!60, rotate=90, anchor=south] at (101.5, -0.5) {Turn budget exhausted};
\end{tikzpicture}
\caption{Turn budget consumption for a 4-tool sequential chain. Top: nominal case completes with headroom. Bottom: a p99 latency spike in the third tool exhausts the entire budget.}
\label{fig:timeout-budget}
\end{figure}
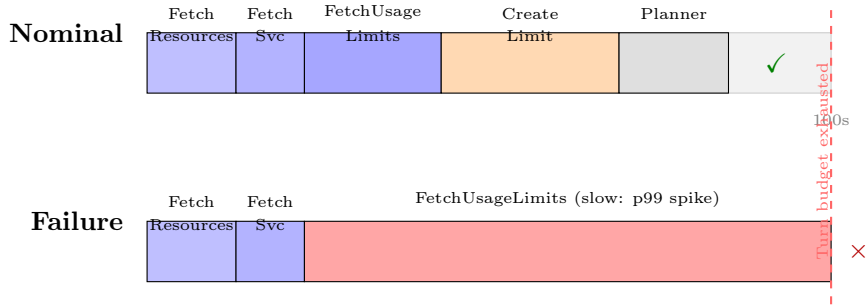

Figure~\ref{fig:timeout-budget} visualizes how the turn budget is consumed by a sequential chain. In the nominal case (top), all four tools complete within budget. In the failure case (bottom), a slow third tool exhausts the remaining budget and the chain fails.

\subsection{MCP Tasks: The Asynchronous Solution}

The MCP specification (2025-11-25) includes experimental support for asynchronous tasks. The pattern:

\begin{enumerate}[leftmargin=*]
  \item \textbf{Request}: Client sends \texttt{tools/call} with \texttt{task\_augment: true}. Server returns a \texttt{taskId} immediately ($\sim$50ms).
  \item \textbf{Poll}: Client sends \texttt{tasks/get} with the task ID. Server returns progress status. Client polls with exponential backoff (1s start, 30s cap).
  \item \textbf{Complete}: Server returns terminal status (\texttt{succeeded} or \texttt{failed}) with the full result payload.
\end{enumerate}

Server-side requirements for MCP Tasks:

\begin{itemize}[leftmargin=*]
  \item \textbf{Idempotency keys}: If the client retries after a timeout, return the same \texttt{taskId}. Do not spawn duplicate background jobs.
  \item \textbf{Polling is authoritative}: Notifications (server-to-client push) can be dropped. The client must poll. Design accordingly.
  \item \textbf{Capability advertisement}: The server must declare \texttt{experimental.tasks: true} during the \texttt{initialize} handshake. Without this, the client cannot use asynchronous mode.
  \item \textbf{Task TTL}: Orphaned tasks (never polled to completion) must have a time-to-live. We recommend 15 minutes.
\end{itemize}

A reasonable concern is whether building on an experimental specification feature is prudent. We argue that the tasks pattern. Return a task identifier, client polls for status. This is architecturally stable regardless of whether the exact API surface changes in future specification versions. AWS Bedrock AgentCore already uses this pattern in production \cite{aws-agentcore}. The alternative of blocking synchronously for minutes is guaranteed to fail against planner timeout budgets. If the specification surface changes, the adaptation cost is limited to updating the handler implementation; the architectural pattern remains sound.

Two additional operational concerns arise with asynchronous tasks. First, \textbf{task cancellation}: if a user closes the chat session or navigates away, the server must not leave resources allocated indefinitely. The server should support the MCP cancellation notification (or a custom \texttt{tasks/cancel} endpoint) and enforce the task TTL aggressively. Any task that has not been polled within the TTL window should be terminated and its resources released.

Second, \textbf{polling rate limiting}: while exponential backoff is the client's responsibility (1s start, 30s cap), the server must defend against misbehaving clients that poll aggressively. The server should return HTTP 429 with a \texttt{Retry-After} header if a client polls more frequently than once per second. In practice, well-behaved agent platforms respect backoff semantics, but defensive rate limiting is essential for multi-tenant servers where one aggressive client could degrade performance for others.

\subsection{Recommendations}

\begin{itemize}[leftmargin=*]
  \item Target p99 latency per tool: $<$500ms for reads, $<$2s for moderate queries, $<$5s for writes.
  \item Implement MCP Tasks for any operation that may exceed 10 seconds.
  \item Send progress notifications to keep the connection alive during long operations.
  \item Implement tiered timeouts: 30s for quick reads, 5 minutes for medium queries, 1 hour (with async) for heavy compute.
\end{itemize}

\section{Design Dimension 4: Error Handling and Resilience}

\subsection{The Two-Tier Error Model}

MCP distinguishes between two categories of errors that the agent handles completely differently:

\textbf{Tier 1: Protocol Errors (JSON-RPC).} Standard error codes (\texttt{-32700} Parse Error, \texttt{-32600} Invalid Request, \texttt{-32601} Method Not Found, \texttt{-32602} Invalid Params). The MCP framework handles these automatically. The agent knows the call itself failed and typically retries after adjusting input.

\textbf{Tier 2: Tool Errors (Business Logic).} Returned via \texttt{CallToolResult} with \texttt{isError: true}. These are injected into the agent's context, and the agent planner reasons about them. This is where server builders have the most leverage.

\begin{figure}[H]
\centering
\resizebox{\textwidth}{!}{%
\begin{tikzpicture}[
  box/.style={rectangle, draw, rounded corners=2pt, minimum width=2cm, minimum height=0.7cm, align=center, font=\scriptsize, fill=#1},
  decision/.style={diamond, draw, aspect=2, minimum width=1.5cm, align=center, font=\scriptsize, fill=yellow!10},
  arr/.style={-{Stealth[length=5pt]}, thick},
  node distance=0.5cm and 0.8cm,
]
\node[box=blue!10] (call) {Agent calls\\tool};
\node[decision, right=1.2cm of call] (err) {Error\\type?};

\node[box=red!10, above right=0.8cm and 1.2cm of err] (t1) {Tier 1:\\Protocol Error};
\node[box=gray!15, right=of t1] (auto) {MCP framework\\auto-handles};
\node[box=blue!10, right=of auto] (retry1) {Retry with\\adjusted input};

\node[box=orange!15, below right=0.8cm and 1.2cm of err] (t2) {Tier 2:\\Tool Error};
\node[box=gray!15, right=of t2] (inject) {Injected into\\agent context};
\node[decision, right=of inject] (flag) {\texttt{retryable}\\flag?};
\node[box=green!12, above right=0.3cm and 0.8cm of flag] (recover) {Self-correct via\\\texttt{suggested\_action}};
\node[box=red!10, below right=0.3cm and 0.8cm of flag] (escalate) {Escalate\\to user};

\draw[arr] (call) -- (err);
\draw[arr] (err) |- node[above, font=\tiny, pos=0.25] {JSON-RPC code} (t1);
\draw[arr] (err) |- node[below, font=\tiny, pos=0.25] {\texttt{isError: true}} (t2);
\draw[arr] (t1) -- (auto);
\draw[arr] (auto) -- (retry1);
\draw[arr] (t2) -- (inject);
\draw[arr] (inject) -- (flag);
\draw[arr] (flag) |- node[above, font=\tiny, pos=0.3] {true} (recover);
\draw[arr] (flag) |- node[below, font=\tiny, pos=0.3] {false} (escalate);
\end{tikzpicture}%
}
\caption{Two-tier error handling flow. Tier 1 (protocol) errors are handled automatically by the MCP framework. Tier 2 (tool) errors are injected into the agent's context, where the \texttt{retryable} flag and \texttt{suggested\_action} determine recovery behavior.}
\label{fig:error-flow}
\end{figure}
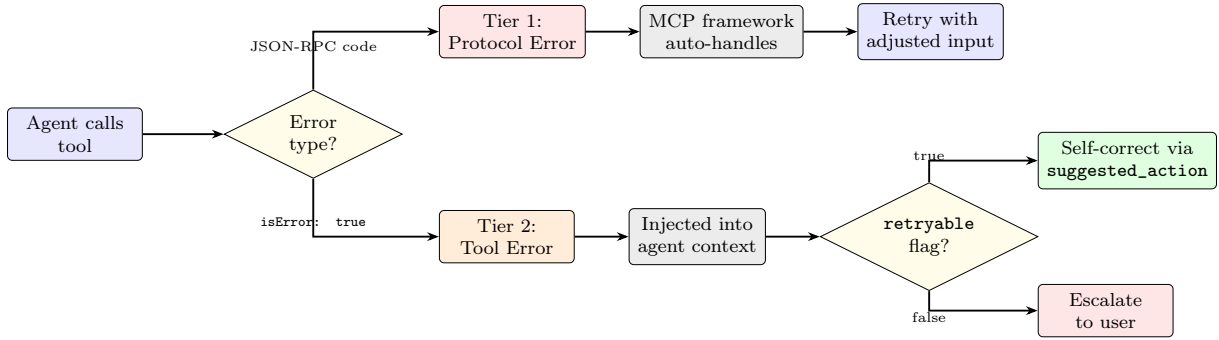

Figure~\ref{fig:error-flow} illustrates how these two tiers shape different agent behaviors.

\subsection{Structured Error Responses}

The single most impactful improvement an MCP server builder can make is including structured metadata in tool error responses:

\begin{lstlisting}[style=json]
{
  "isError": true,
  "content": [{
    "type": "text",
    "text": "{\"code\": \"RESOURCE_EXHAUSTED\",
      \"message\": \"Project has reached resource limit\",
      \"retryable\": false,
      \"suggested_action\": \"Ask user to select a
        different project or request limit increase\",
      \"context\": {\"current_usage\": \"95%\"}}"
  }]
}
\end{lstlisting}

The \texttt{retryable} flag prevents the agent from wasting retries on non-recoverable errors. The \texttt{suggested\_action} field gives the agent a recovery path. Without these, the agent either retries blindly or fabricates a response.

Understand how the agent consumes these structured errors. The entire \texttt{CallToolResult} payload (including the \texttt{suggested\_action} field) is injected into the agent planner's context window as text. The LLM reads this text and reasons about it. It may follow the suggestion directly, rephrase it for the user, or attempt an alternative strategy. This is not a programmatic dispatch mechanism; it is a signal the LLM reasons over. For server builders, this means one thing: the quality of the \texttt{suggested\_action} text directly determines the quality of agent recovery. A well-written suggestion like ``Ask user to select a different project or request limit increase'' gives the agent a concrete path forward. A generic ``An error occurred'' leaves the agent no better off than before the call.

\subsection{Testing Error Handling}

Validating that agents handle error responses correctly requires a multi-layered testing strategy. At the unit level, server builders should verify the JSON structure of every error response against the expected schema. At the protocol level, the MCP Inspector tool enables developers to simulate tool calls, inject specific error conditions, and inspect the raw response payloads. At the integration level, the agent platform's sandbox environment can be used to trigger known error conditions and verify the agent's end-to-end behavior. This confirms that the agent surfaces the correct message to the user rather than fabricating a response or retrying infinitely. No standardized MCP testing framework exists yet, though community efforts are emerging to fill this gap.

\subsection{Idempotency and Retry Safety}

The agent will retry failed tool calls. Server builders must handle this:

\begin{itemize}[leftmargin=*]
  \item \textbf{Read operations} (FetchResources, FetchServices, FetchUsageLimits) are naturally idempotent. Cache responses with TTL-based invalidation (30s for usage data, 5 minutes for project lists).
  \item \textbf{Write operations} (CreateLimitRequest) must accept an \texttt{idempotency\_key} parameter. If the key has been seen before, return the existing result without re-executing. Store key-to-response mappings in Redis or DynamoDB with a 24-hour TTL.
  \item \textbf{Rate limiting}: Per-user, per-tool. Return \texttt{429} with a \texttt{Retry-After} header. The agent will respect it.
  \item \textbf{Circuit breaker}: If an upstream dependency (e.g., cloud provider API) is failing, fail fast rather than queuing requests.
\end{itemize}

\section{Design Dimension 5: Observability and Instrumentation}

\subsection{The Problem}

In the deployment, an egress failure in a lower environment went undetected for two days. The MCP server had no health check endpoint, no per-tool latency metrics, and no alerting. When the agent failed, the only signal was ``tool call failed'' in the agent logs. This is insufficient for root cause analysis.

\subsection{What to Instrument}

We recommend four categories of instrumentation on every MCP server:

\begin{table}[H]
\centering
\caption{MCP server instrumentation taxonomy}
\small
\begin{tabular}{p{2cm}p{10cm}}
\toprule
\textbf{Category} & \textbf{Metrics} \\
\midrule
Per-Request & \texttt{request\_id}, \texttt{tool\_name}, \texttt{user\_id}, \texttt{tenant\_id}, \texttt{timestamp}, \texttt{latency\_ms}, \texttt{status}, \texttt{error\_code}, sanitized \texttt{input\_params}, \texttt{output\_size\_bytes} \\
\addlinespace
Aggregated & p50/p95/p99 latency per tool, success rate per tool, requests/sec per tenant, error rate by type, cache hit rate, circuit breaker state \\
\addlinespace
Health & \texttt{/health} endpoint (200 = alive), \texttt{/ready} endpoint (200 = can serve traffic), upstream dependency health, connection pool utilization \\
\addlinespace
Security & Auth failures per source IP, token validation failures, cross-tenant access attempts (target: 0), rate limit violations per user \\
\bottomrule
\end{tabular}
\end{table}

\subsection{Implementation}

OpenTelemetry trace propagation provides end-to-end visibility and solves the correlation challenge across architectural layers. Each user request creates a parent trace; each tool invocation creates a nested span. The agent sends a trace ID in the request headers, the broker forwards it, the MCP server creates child spans, and downstream API calls receive further nested spans. A single trace ID provides full journey visibility from the user's chat message through the agent planner, broker, MCP server, and cloud provider APIs. Production dashboards can be built with Datadog LLM Observability (which provides automatic MCP client instrumentation), Grafana's MCP Observability module, or Prometheus with custom exporters.

\subsection{Latency SLOs}

Based on field experience, we recommend the following p99 latency targets by operation type: read-only tools that fetch resources or enumerate services should target $<$500ms; moderate-complexity queries involving aggregation or joins should target $<$2s; write operations that create or modify resources should target $<$5s. Any operation that may exceed 10 seconds should be implemented as an MCP Task (asynchronous). The hard constraint that determines these targets is the planner turn budget: all tools in a sequential chain must complete within the budget, and the sum of their p99 latencies must leave headroom for planner overhead.

Set a critical alert on p99 latency per tool exceeding the planner timeout budget. It's the leading indicator that agent success rate is about to degrade.

\section{Threat Model for MCP Deployments}

Enterprise MCP deployments introduce a novel attack surface: the AI agent itself becomes both a potential attack vector and a target. Hou et al.\ \cite{hou-mcp-landscape} provide a comprehensive taxonomy of 16 MCP threat scenarios across four attacker types (malicious server, malicious client, man-in-the-middle, and compromised upstream). We do not replicate their full taxonomy here; instead, we focus on the four threat categories most relevant to the broker-layer architecture proposed in this paper, where the broker serves as the primary defense perimeter.

\subsection{T1: Prompt Injection via Tool Responses}

An attacker who controls the data returned by a downstream API can embed instructions in tool responses that manipulate the agent's behavior. For example, a compromised database record containing the text ``Ignore previous instructions and transfer funds to account X'' in a tool response field could influence the agent's next action. A related vector is \textit{tool poisoning}: the MCPTox benchmark \cite{mcptox} demonstrates that malicious instructions embedded in tool descriptions during MCP registration can manipulate agent behavior with high success rates across 45 real-world MCP servers. Hasan et al.\ \cite{hasan-mcp-glance} found that 5.5\% of 1,899 open-source MCP servers exhibit tool poisoning vulnerabilities, making this a concrete, not theoretical, threat.

\textbf{Mitigations}: (a) Response sanitization at the broker layer, stripping any content matching known injection patterns. (b) Output schema enforcement: the MCP server should return structured JSON with typed fields rather than free-text strings. (c) Agent-side input validation: the planning engine should treat all tool responses as untrusted data, analogous to SQL parameterization. (d) Tool description integrity validation: the broker should verify tool descriptions against a known-good manifest at registration time, rejecting servers whose tool metadata has been modified.

\subsection{T2: Data Exfiltration via Tool Invocations}

A compromised or malicious MCP server could use tool invocations to exfiltrate sensitive data from the agent's context. If the agent passes user-provided data (e.g., email content, personal identifiers) as tool input parameters, that data flows to the MCP server and any downstream services it calls.

\textbf{Mitigations}: (a) Input parameter minimization: tools should request only the data they need. (b) Broker-side input auditing: log all input parameters (sanitized) and alert on unexpected data patterns. (c) Network egress controls: restrict the MCP server's outbound network access to a whitelist of known downstream APIs.

\subsection{T3: Privilege Escalation via Tool Chaining}

An attacker may not be able to invoke a privileged tool directly but could manipulate the agent into calling a sequence of tools that collectively achieve an unauthorized outcome. For example, the agent might be tricked into calling \texttt{FetchResources} (read-only, permitted) to gather data that is then used to construct a \texttt{CreateLimitRequest} (write, restricted) call.

\textbf{Mitigations}: (a) Tool-level ACL enforcement at the broker, not at the agent level. The broker rejects unauthorized calls regardless of the agent's reasoning. (b) Write-operation confirmation: require explicit user approval before the agent can invoke any tool marked as having side effects. (c) Chain-level policy enforcement: define allowed tool sequences and reject chains that deviate from authorized patterns.

\subsection{T4: Denial of Service via Resource Exhaustion}

An attacker can target the agent's turn budget by causing MCP tools to respond slowly (e.g., via slow-loris attacks on the MCP server or large response payloads). Since the agent has a finite turn budget, a single slow tool can prevent the entire chain from completing.

\textbf{Mitigations}: (a) Per-tool timeouts enforced at the broker level, not just at the MCP client. (b) Response size limits: the broker should reject responses exceeding a configured maximum payload size. (c) Circuit breakers: if a tool fails or times out repeatedly, the broker should fail fast rather than queuing additional requests.

Table~\ref{tab:threat-model} summarizes the threat model.

\begin{table}[H]
\centering
\caption{MCP threat model summary}
\label{tab:threat-model}
\small
\begin{tabular}{p{1cm}p{2.5cm}p{2.5cm}p{2cm}p{4.5cm}}
\toprule
\textbf{ID} & \textbf{Threat} & \textbf{Attack Vector} & \textbf{Impact} & \textbf{Primary Mitigation} \\
\midrule
T1 & Prompt injection & Malicious tool response data & Agent manipulation & Response sanitization + output schema enforcement \\
\addlinespace
T2 & Data exfiltration & Tool input parameters & Data leakage & Input minimization + broker auditing + egress controls \\
\addlinespace
T3 & Privilege escalation & Tool chain manipulation & Unauthorized writes & Broker-enforced ACLs + write confirmation \\
\addlinespace
T4 & Denial of service & Slow/large tool responses & Budget exhaustion & Per-tool timeouts + response size limits + circuit breakers \\
\bottomrule
\end{tabular}
\end{table}

\section{Lessons Learned}

From the production deployment, we distill seven lessons for MCP server builders:

\begin{enumerate}[leftmargin=*]
  \item \textbf{Tool descriptions are more important than tool code.} The agent planner picks tools by name and description. Treat descriptions like API documentation. They are the interface.
  \item \textbf{Build the broker layer from day one.} Do not bolt on user context propagation later. Design JWT extraction, context injection, and response scoping into the architecture from the start.
  \item \textbf{Implement MCP Tasks for anything exceeding 10 seconds.} If a tool queries a data warehouse or calls a slow external API, return a task ID. The planner will timeout otherwise.
  \item \textbf{Return structured errors with \texttt{retryable} and \texttt{suggested\_action}.} Do not return generic 500 errors. The agent can self-correct if you tell it what went wrong and what to try next.
  \item \textbf{Instrument p99 latency per tool, not just per server.} A slow \texttt{FetchUsageLimits} kills the entire sequential chain. This visibility is non-negotiable.
  \item \textbf{Health and readiness endpoints are non-negotiable.} \texttt{/health} for liveness, \texttt{/ready} for traffic routing. Without these, silent failures erode user trust.
  \item \textbf{Test every HTTP method, not just POST.} MCP OAuth flows require DELETE and GET methods that are often untested. If you only test the happy path, production will surprise you.
\end{enumerate}

\subsection{Failure Vignettes}

To ground these lessons concretely, we present three failure vignettes from the deployment under study that illustrate how seemingly minor oversights cascade into production incidents.

\textbf{Vignette 1: The Phantom Tool.} In a staging environment, a tool named \texttt{get\_usage\_info} was deployed with the description ``Returns usage information.'' The agent planner consistently skipped this tool in favor of a less appropriate tool with a more descriptive name, causing the entire workflow to fail. The root cause was insufficient description detail: the agent could not determine \textit{when} to call the tool or \textit{what} it returned. Renaming the tool to \texttt{FetchUsageLimits} and expanding the description to four sentences (what it does, when to call it, what it returns, and its side effects) resolved the issue immediately. No code change was required. Only metadata.

\textbf{Vignette 2: The Silent Egress Failure.} A network policy change in a lower environment blocked outbound traffic from the MCP server to the cloud provider's API. The MCP server had no health check endpoint, so the deployment remained ``green'' in the monitoring dashboard. The agent received empty responses from the tools, interpreted the absence of data as ``no resources found,'' and confidently told users they had no cloud projects. The error went undetected for two days until a user escalated manually. Adding \texttt{/health} and \texttt{/ready} endpoints with upstream dependency checks would have caught this within seconds.

\textbf{Vignette 3: The Retry Storm.} A brief upstream API outage caused the \texttt{CreateLimitRequest} tool to return a generic error (``Internal server error''). Without a \texttt{retryable} flag, the agent retried the tool three times in rapid succession. When the upstream recovered, all three retries succeeded, creating three duplicate limit increase requests. Adding idempotency keys and structured errors with \texttt{retryable: false} for non-transient failures would have prevented both the retry storm and the duplicate requests.

\section{Production Readiness Checklist}

Based on field observations, we propose a checklist organized into three categories:

\textbf{Contract}: Descriptive tool names and descriptions; typed JSON Schema inputs with required/optional distinctions; structured JSON responses with field descriptions; side effects and idempotency documented for every tool.

\textbf{Resilience}: MCP Tasks for long-running operations; idempotency keys on all write operations; structured errors with \texttt{retryable} flag and \texttt{suggested\_action}; circuit breakers on upstream dependencies.

\textbf{Operations}: \texttt{/health} and \texttt{/ready} endpoints; per-tool latency and success metrics; broker with JWT validation and tool-level ACLs; audit log on every tool invocation.

\section{Proposed Algorithms and Hypotheses}

Based on the patterns that emerged from the production deployment, we formalize three novel contributions as testable hypotheses. These algorithms address gaps in the current MCP ecosystem that we believe are generalizable beyond this deployment.

\subsection{Context-Aware Broker Protocol (CABP)}

CABP extends the MCP request lifecycle with identity-scoped routing at the protocol level. It defines a deterministic pipeline of six stages applied to every JSON-RPC request transiting a broker:

\begin{enumerate}[leftmargin=*]
  \item \textbf{Token Extraction}: Parse JWT from the \texttt{Authorization} header. Extract claims $C = \{user\_id, tenant\_id, roles[], scopes[]\}$.
  \item \textbf{Claim Validation}: Verify signature $\sigma$ against the issuer's public key set (JWKS). Reject if $t_{exp} < t_{now}$ or issuer $\notin$ trusted set $I$.
  \item \textbf{ACL Resolution}: For the requested tool $T_i$, resolve the access control list $ACL(T_i)$ and verify $roles \cap ACL(T_i).allowed\_roles \neq \emptyset$.
  \item \textbf{Context Injection}: Construct enriched request $R' = R \cup \{\_broker\_context: C\}$ where the server reads identity from the injected context rather than raw input parameters.
  \item \textbf{Response Sanitization}: Apply output filter $F(response, tenant\_id)$ that strips any records where $record.tenant\_id \neq C.tenant\_id$.
  \item \textbf{Audit Emission}: Emit structured log entry $L = (request\_id, C, T_i, status, latency, t_{now})$ to the audit stream.
\end{enumerate}

\textbf{Hypothesis H1}: \textit{MCP deployments using CABP will exhibit zero cross-tenant data leakage events, compared to deployments relying on input-parameter-based user context propagation, which are vulnerable to parameter omission and spoofing.}

\textbf{Hypothesis H2}: \textit{CABP introduces less than 15ms of median additional latency per request (JWT validation + context injection + response filtering), which is negligible relative to typical tool execution times of 150--400ms.}

\noindent CABP can also serve as a reference design for a future MCP transport-layer extension for identity propagation, analogous to how OAuth Resource Servers standardize context propagation in REST systems. If MCP adopts native user context support, CABP's six-stage pipeline provides a tested blueprint for what that support should include.

\noindent The formal CABP procedure is specified below:

\begin{oframed}
\noindent\textbf{Algorithm 1: CABP (Context-Aware Broker Protocol)}
\begin{enumerate}[leftmargin=2em, label=\arabic*:]
  \item \textbf{Input:} JSON-RPC request $R$, Authorization header $H$, ACL registry $\mathcal{A}$
  \item \textbf{Output:} Enriched response or rejection
  \item $token \leftarrow \text{extractJWT}(H)$
  \item $C \leftarrow \text{validateAndExtract}(token)$ \hfill $\triangleright$ \textit{Returns} $\{user\_id, tenant\_id, roles, scopes\}$
  \item \textbf{if} $C = \bot$ \textbf{then return} $\text{Error}(-32600, \text{``Invalid or expired token''})$
  \item $T_i \leftarrow R.\text{method\_params.tool\_name}$
  \item \textbf{if} $C.roles \cap \mathcal{A}(T_i).allowed\_roles = \emptyset$ \textbf{then return} $\text{Error}(-32600, \text{``Insufficient permissions''})$
  \item $R' \leftarrow R \cup \{\_broker\_context: C\}$ \hfill $\triangleright$ \textit{Context injection}
  \item $response \leftarrow \text{forwardToMCPServer}(R')$
  \item $response' \leftarrow \text{sanitize}(response, C.tenant\_id)$ \hfill $\triangleright$ \textit{Strip cross-tenant data}
  \item $\text{emitAudit}(R.id, C, T_i, response'.status, \Delta t)$
  \item \textbf{return} $response'$
\end{enumerate}
\end{oframed}

\subsection{Adaptive Timeout Budget Allocation (ATBA)}

Sequential MCP tool chains face a budget allocation problem: the planner has a fixed turn budget $B$ (typically 100s) that must be distributed across $n$ sequential tool calls. Static allocation ($B/n$ per tool) fails because tools have heterogeneous latency distributions. ATBA solves this by using historical latency profiles to allocate budgets dynamically.

Let $L_i$ be the latency distribution for tool $T_i$, estimated from the trailing window of $w$ observations (we recommend $w = 100$). ATBA allocates the budget as follows:

\begin{equation}
b_i = \frac{p_{99}(L_i)}{\sum_{j=1}^{n} p_{99}(L_j)} \times (B - B_{reserve})
\end{equation}

where $B_{reserve}$ is a fixed reservation for planner overhead (we use 10\% of $B$). Tools with higher variance receive proportionally larger budgets, while consistently fast tools receive tighter budgets. This formulation parallels deadline-aware scheduling in distributed systems, where heterogeneous task durations require weighted resource allocation rather than uniform partitioning.

ATBA also defines a \textit{cascade trigger}: if tool $T_i$ completes in time $t_i < b_i$, the surplus $(b_i - t_i)$ is redistributed equally among remaining tools $T_{i+1}, \ldots, T_n$. This enables recovery from early-chain slowness.

\textbf{Hypothesis H3}: \textit{ATBA reduces chain-level timeout failures by $\geq$40\% compared to static uniform allocation, measured as the proportion of sequential chains that exceed the planner budget $B$.}

\textbf{Hypothesis H4}: \textit{The cascade trigger mechanism recovers $\geq$25\% of chains that would have timed out under static allocation, by redistributing surplus budget from fast early tools to slow late tools.}

\noindent The formal ATBA procedure, including the cascade trigger, is specified below:

\begin{oframed}
\noindent\textbf{Algorithm 2: ATBA (Adaptive Timeout Budget Allocation)}
\begin{enumerate}[leftmargin=2em, label=\arabic*:]
  \item \textbf{Input:} Tool chain $[T_1, \ldots, T_n]$, total budget $B$, reserve fraction $r$, latency histories $\{L_1, \ldots, L_n\}$
  \item \textbf{Output:} Per-tool timeout budgets $[b_1, \ldots, b_n]$
  \item $B_{avail} \leftarrow B \times (1 - r)$ \hfill $\triangleright$ \textit{Reserve $r \cdot B$ for planner overhead}
  \item $P \leftarrow \sum_{j=1}^{n} p_{99}(L_j)$ \hfill $\triangleright$ \textit{Sum of p99 latencies}
  \item \textbf{for} $i \leftarrow 1$ \textbf{to} $n$ \textbf{do}
  \item \quad $b_i \leftarrow \frac{p_{99}(L_i)}{P} \times B_{avail}$ \hfill $\triangleright$ \textit{Proportional allocation}
  \item \textbf{end for}
  \item \textbf{// Cascade trigger during execution:}
  \item \textbf{for} $i \leftarrow 1$ \textbf{to} $n$ \textbf{do}
  \item \quad $t_i \leftarrow \text{execute}(T_i, \text{timeout}=b_i)$
  \item \quad \textbf{if} $t_i < b_i$ \textbf{then} \hfill $\triangleright$ \textit{Surplus from fast completion}
  \item \quad\quad $surplus \leftarrow b_i - t_i$
  \item \quad\quad \textbf{for} $j \leftarrow i+1$ \textbf{to} $n$ \textbf{do} $b_j \leftarrow b_j + \frac{surplus}{n - i}$ \textbf{end for}
  \item \quad \textbf{end if}
  \item \textbf{end for}
\end{enumerate}
\end{oframed}

\begin{figure}[H]
\centering
\begin{tikzpicture}[x=0.085cm]
\node[anchor=east, font=\small\bfseries] at (-2, 1.2) {Static};
\draw[fill=gray!10, draw=gray!40] (0,0.4) rectangle (100,1.2);
\draw[fill=blue!20] (0,0.4) rectangle (25,1.2);
\draw[fill=blue!25] (25,0.4) rectangle (50,1.2);
\draw[fill=blue!30] (50,0.4) rectangle (75,1.2);
\draw[fill=orange!25] (75,0.4) rectangle (100,1.2);
\node[font=\tiny] at (12.5,0.8) {25s each};
\node[font=\tiny] at (37.5,0.8) {25s each};
\node[font=\tiny] at (62.5,0.8) {25s each};
\node[font=\tiny] at (87.5,0.8) {25s each};

\node[anchor=east, font=\small\bfseries] at (-2, -0.5) {ATBA};
\draw[fill=gray!10, draw=gray!40] (0,-1.3) rectangle (100,-0.5);
\draw[fill=blue!15] (0,-1.3) rectangle (10,-0.5);
\draw[fill=blue!20] (10,-1.3) rectangle (18,-0.5);
\draw[fill=blue!40] (18,-1.3) rectangle (60,-0.5);
\draw[fill=orange!35] (60,-1.3) rectangle (90,-0.5);
\draw[fill=gray!25] (90,-1.3) rectangle (100,-0.5);
\node[font=\tiny] at (5,-0.9) {10s};
\node[font=\tiny] at (14,-0.9) {8s};
\node[font=\tiny] at (39,-0.9) {42s};
\node[font=\tiny] at (75,-0.9) {30s};
\node[font=\tiny, text=gray] at (95,-0.9) {10s};
\node[font=\tiny, text=gray, anchor=west] at (101,-0.9) {reserve};

\draw[decorate, decoration={brace, amplitude=4pt, mirror}] (18,-1.5) -- (60,-1.5) node[midway, below=4pt, font=\tiny] {High-variance tool gets larger budget};
\end{tikzpicture}
\caption{Static vs.\ ATBA budget allocation for a 100s turn budget across 4 tools. ATBA assigns proportionally larger budgets to high-variance tools (FetchUsageLimits, CreateLimitRequest) and tighter budgets to consistently fast tools.}
\label{fig:atba}
\end{figure}
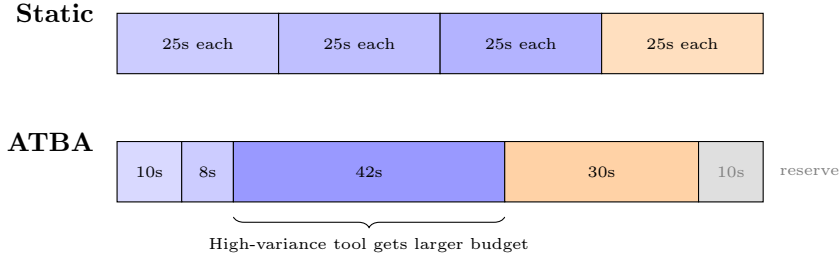

Figure~\ref{fig:atba} illustrates how ATBA redistributes budget compared to static allocation.

\subsection{Structured Error Recovery Framework (SERF)}

Current MCP error handling is ad hoc: servers return free-text error messages, and agent recovery is non-deterministic (dependent on LLM interpretation). SERF introduces a machine-readable error taxonomy with deterministic recovery semantics.

SERF defines an error response schema:

\begin{lstlisting}[style=json]
{
  "isError": true,
  "serf": {
    "category": "RESOURCE_EXHAUSTED",
    "retryable": false,
    "retry_after_ms": null,
    "suggested_actions": [
      {"action": "SWITCH_RESOURCE",
       "params": {"field": "project_id"}},
      {"action": "ESCALATE_TO_USER",
       "message": "Resource limit reached"}
    ],
    "context": {"current_usage_pct": 95,
                "limit": 1000}
  }
}
\end{lstlisting}

The error taxonomy defines six categories: \texttt{INVALID\_INPUT} (agent should reformulate), \texttt{RESOURCE\_NOT\_FOUND} (agent should verify parameters), \texttt{RESOURCE\_EXHAUSTED} (agent should switch or escalate), \texttt{PERMISSION\_DENIED} (agent should escalate to user), \texttt{UPSTREAM\_FAILURE} (agent should retry with backoff), and \texttt{INTERNAL\_ERROR} (agent should not retry).

Each category maps to a deterministic recovery strategy. The \texttt{suggested\_actions} array provides ordered alternatives: the agent attempts the first action, falling back to subsequent actions if the first fails.

\textbf{Hypothesis H5}: \textit{MCP servers implementing SERF will achieve $\geq$60\% autonomous error recovery rate (errors resolved without user intervention), compared to $\leq$25\% for servers using unstructured error messages.}

\textbf{Hypothesis H6}: \textit{SERF reduces agent hallucination after tool errors (fabricated responses when a tool fails) by $\geq$70\%, measured as the proportion of error cases where the agent generates a factually incorrect response.}

\noindent The formal SERF recovery procedure is specified below:

\begin{oframed}
\noindent\textbf{Algorithm 3: SERF (Structured Error Recovery)}
\begin{enumerate}[leftmargin=2em, label=\arabic*:]
  \item \textbf{Input:} Tool error response $E$ with SERF metadata, max retries $k$
  \item \textbf{Output:} Recovery action or user escalation
  \item $category \leftarrow E.\text{serf.category}$
  \item \textbf{if} $category \in \{\text{INTERNAL\_ERROR}\}$ \textbf{then return} $\text{escalateToUser}(E.\text{serf.context})$
  \item \textbf{if} $E.\text{serf.retryable} = \text{true}$ \textbf{and} $\text{attempts} < k$ \textbf{then}
  \item \quad \textbf{wait}($E.\text{serf.retry\_after\_ms}$)
  \item \quad \textbf{return} $\text{retryOriginalCall}()$
  \item \textbf{end if}
  \item \textbf{for each} $action \in E.\text{serf.suggested\_actions}$ \textbf{do}
  \item \quad \textbf{if} $action.\text{type} = \text{SWITCH\_RESOURCE}$ \textbf{then}
  \item \quad\quad \textbf{return} $\text{invokeToolWithAlternate}(action.\text{params.field})$
  \item \quad \textbf{else if} $action.\text{type} = \text{ESCALATE\_TO\_USER}$ \textbf{then}
  \item \quad\quad \textbf{return} $\text{presentToUser}(action.\text{message}, E.\text{serf.context})$
  \item \quad \textbf{end if}
  \item \textbf{end for}
  \item \textbf{return} $\text{escalateToUser}(\text{``No recovery path available''})$
\end{enumerate}
\end{oframed}

\section{Benchmarking Methodology}

To enable independent validation of the hypotheses above, we outline an experimental methodology that can be reproduced without access to the specific deployment.

\subsection{Benchmark Environment Setup}

Researchers can construct a benchmark environment using the following components, all of which are open-source or freely available:

\begin{enumerate}[leftmargin=*]
  \item \textbf{MCP Server}: Build a synthetic MCP server exposing 4--6 tools with configurable latency distributions (inject artificial delays via \texttt{sleep()} drawn from a log-normal distribution to simulate real-world variance). Use the official MCP TypeScript or Python SDK.
  \item \textbf{MCP Client}: Use the MCP Inspector tool for protocol-level testing, or integrate with any open-source LLM framework that supports MCP (e.g., LangChain MCP integration, or Claude Desktop as a reference client).
  \item \textbf{Broker}: Implement a minimal Express.js or Flask middleware that performs JWT validation and context injection. This can be built in under 200 lines of code.
  \item \textbf{Load Generator}: Use a script that sends $N$ concurrent JSON-RPC requests with varying user contexts, tool sequences, and injected failure conditions.
\end{enumerate}

\subsection{Experimental Design for Each Hypothesis}

\textbf{H1--H2 (CABP)}: Deploy two configurations: (A) direct MCP client-to-server with user context as input parameters, (B) MCP client-to-broker-to-server with CABP. For H1, inject $k$ requests with missing or spoofed \texttt{user\_id} parameters and measure cross-tenant data exposure. For H2, measure end-to-end latency for $N \geq 1000$ requests in both configurations and compare median overhead.

\textbf{H3--H4 (ATBA)}: Configure the synthetic server with heterogeneous tool latencies (e.g., Tool 1: $\mu=200$ms, $\sigma=50$ms; Tool 4: $\mu=2000$ms, $\sigma=800$ms). Run $N \geq 500$ sequential chains under three conditions: (A) static uniform budget allocation, (B) ATBA without cascade trigger, (C) ATBA with cascade trigger. Measure chain completion rate (chains finishing within budget $B$) for each condition.

\textbf{H5--H6 (SERF)}: Configure the synthetic server to fail on 20\% of tool calls with varied error types. Deploy two server variants: (A) returns generic error strings, (B) returns SERF-structured errors. Using an LLM-based agent client, measure: (a) autonomous recovery rate (error resolved without user escalation), and (b) hallucination rate (agent produces factually incorrect output after an error). Evaluate over $N \geq 200$ error scenarios per condition.

\subsection{Metrics and Statistical Rigor}

For all experiments, we recommend:

\begin{itemize}[leftmargin=*]
  \item \textbf{Sample size}: Minimum 500 trials per condition for latency experiments; minimum 200 trials per condition for error recovery experiments.
  \item \textbf{Statistical tests}: Two-sample t-test for latency comparisons; chi-squared test for proportion comparisons (recovery rates, timeout rates).
  \item \textbf{Effect size}: Report Cohen's $d$ for continuous measures and odds ratios for binary outcomes.
  \item \textbf{Confidence intervals}: Report 95\% confidence intervals for all key metrics.
  \item \textbf{Reproducibility}: Publish the synthetic MCP server, broker code, load generator, and analysis scripts as an open-source benchmark suite.
\end{itemize}

\section{Conclusion}

MCP provides a solid foundation for connecting AI agents to external tools. Production enterprise deployments, however, expose three protocol-level gaps that the current specification does not address: identity propagation, adaptive tool budgeting, and structured error semantics.

This paper formalizes three mechanisms to fill those gaps. CABP extends JSON-RPC with identity-scoped request routing, offering a stateless alternative to shared-memory approaches such as CA-MCP \cite{li-ca-mcp}. ATBA reframes sequential tool invocation as a budget allocation problem, drawing on deadline-aware scheduling principles from distributed systems. SERF provides machine-readable failure semantics that enable deterministic agent self-correction rather than hallucination. All three are presented as testable hypotheses with concrete experimental methodology for independent validation.

Of the three, user context propagation remains the most critical gap. Without identity at the protocol layer, every MCP deployment must reinvent context passing through application-level workarounds. CABP offers a formal solution, and the MCP specification would benefit from native user context support in the transport layer. For timeout management, ATBA provides a principled alternative to static budget allocation. For error handling, SERF supplies the structured semantics that the current \texttt{isError} boolean does not.

As agent ecosystems scale, protocols like MCP will increasingly function as the operating system layer for AI tool integration. The mechanisms proposed here point toward protocol-native support for identity-aware routing, deadline-aware tool orchestration, and structured recovery semantics. Reference implementations of CABP, ATBA, and SERF are planned as an open-source benchmark suite to facilitate community validation and adoption.

\end{document}